\documentclass[10pt,a4paper,onecolumn]{article}
\usepackage{marginnote}
\usepackage{graphicx}
\usepackage{xcolor}
\usepackage{authblk,etoolbox}
\usepackage{titlesec}
\usepackage{calc}
\usepackage{tikz}
\usepackage{hyperref}
\hypersetup{colorlinks,breaklinks,
            colorlinks=[rgb]{0.0, 0.5, 1.0}, 
            citecolor=[rgb]{0.0, 0.5, 1.0},
            urlcolor=[rgb]{0.0, 0.5, 1.0},
            linkcolor=[rgb]{0.0, 0.5, 1.0}}
\usepackage{caption}
\usepackage{amssymb,amsmath}
\usepackage{ifxetex,ifluatex}
\usepackage{fixltx2e} 

\usepackage{natbib}

\usepackage[top=3.5cm, bottom=3cm, right=1.5cm, left=1.0cm,
            headheight=2.2cm, reversemp, includemp, marginparwidth=4.5cm]{geometry}

\titleformat{\section}
  {\normalfont\sffamily\Large\bfseries}
  {}{0pt}{}
\titleformat{\subsection}
  {\normalfont\sffamily\large\bfseries}
  {}{0pt}{}
\titleformat{\subsubsection}
  {\normalfont\sffamily\bfseries}
  {}{0pt}{}
\titleformat*{\paragraph}
  {\sffamily\normalsize}

\usepackage{fancyhdr}
\makeatletter
\let\ps@plain\ps@fancy
\fancyheadoffset[L]{4.5cm}
\fancyfootoffset[L]{4.5cm}

\definecolor{linky}{rgb}{0.0, 0.5, 1.0}

\newcommand{\ExternalLink}{%
   \tikz[x=1.2ex, y=1.2ex, baseline=-0.05ex]{%
       \begin{scope}[x=1ex, y=1ex]
           \clip (-0.1,-0.1)
               --++ (-0, 1.2)
               --++ (0.6, 0)
               --++ (0, -0.6)
               --++ (0.6, 0)
               --++ (0, -1);
           \path[draw,
               line width = 0.5,
               rounded corners=0.5]
               (0,0) rectangle (1,1);
       \end{scope}
       \path[draw, line width = 0.5] (0.5, 0.5)
           -- (1, 1);
       \path[draw, line width = 0.5] (0.6, 1)
           -- (1, 1) -- (1, 0.6);
       }
   }

\patchcmd{\@maketitle}{center}{flushleft}{}{}
\patchcmd{\@maketitle}{center}{flushleft}{}{}
\patchcmd{\@maketitle}{\LARGE}{\LARGE\sffamily}{}{}
\def\maketitle{{%
  
  \AB@maketitle}}
\makeatletter
\renewcommand\AB@affilsepx{ \protect\Affilfont}
\renewcommand\AB@affilnote[1]{{\bfseries #1}\hspace{3pt}}
\makeatother

\renewcommand\Affilfont{\sffamily\small\mdseries}
\ifnum 0\ifxetex 1\fi\ifluatex 1\fi=0 
  \usepackage[T1]{fontenc}
  \usepackage[utf8]{inputenc}

\else 
  \ifxetex
    \usepackage{mathspec}
  \else
    \usepackage{fontspec}
  \fi
  \defaultfontfeatures{Ligatures=TeX,Scale=MatchLowercase}

\fi
\IfFileExists{upquote.sty}{\usepackage{upquote}}{}
\IfFileExists{microtype.sty}{%
\usepackage{microtype}
\UseMicrotypeSet[protrusion]{basicmath} 
}{}

\usepackage{hyperref}
\hypersetup{unicode=true,
            pdftitle={pyEQUIB Python Package, an addendum to proEQUIB: IDL Library for Plasma Diagnostics and Abundance Analysis},
            pdfborder={0 0 0},
            breaklinks=true}
\usepackage{graphicx,grffile}
\makeatletter
\def\maxwidth{\ifdim\Gin@nat@width>\linewidth\linewidth\else\Gin@nat@width\fi}
\def\maxheight{\ifdim\Gin@nat@height>\textheight\textheight\else\Gin@nat@height\fi}
\makeatother
\setkeys{Gin}{width=\maxwidth,height=\maxheight,keepaspectratio}
\IfFileExists{parskip.sty}{%
\usepackage{parskip}
}{
\setlength{\parindent}{0pt}
\setlength{\parskip}{6pt plus 2pt minus 1pt}
}
\ifx\paragraph\undefined\else
\let\oldparagraph\paragraph
\renewcommand{\paragraph}[1]{\oldparagraph{#1}\mbox{}}
\fi
\ifx\subparagraph\undefined\else
\let\oldsubparagraph\subparagraph
\renewcommand{\subparagraph}[1]{\oldsubparagraph{#1}\mbox{}}
\fi

\title{pyEQUIB Python Package, an addendum to proEQUIB: IDL Library for Plasma Diagnostics and Abundance Analysis}

        \author[1, 2, 3]{Ashkbiz Danehkar}
    
      \affil[1]{Department of Physics and Astronomy, Macquarie University, Sydney, NSW 2109, Australia}
      \affil[2]{Harvard-Smithsonian Center for Astrophysics, 60 Garden Street, Cambridge, MA 02138, USA}
      \affil[3]{Department of Astronomy, University of Michigan, 1085 S. University Avenue, Ann Arbor, MI 48109, USA}
  \date{\vspace{-5ex}}

\begin{document}
\maketitle

\marginpar{
  \sffamily\small

  {\bfseries DOI:} \href{https://doi.org/10.21105/joss.02798}{\color{linky}{10.21105/joss.02798}}

  \vspace{2mm}

  {\bfseries Software}
  \begin{itemize}
    \setlength\itemsep{0em}
    \item \href{https://github.com/openjournals/joss-reviews/issues/2798}{\color{linky}{Review}} \ExternalLink
    \item \href{https://github.com/equib/pyEQUIB}{\color{linky}{Repository}} \ExternalLink
    \item \href{https://doi.org/10.5281/zenodo.4287576}{\color{linky}{Archive}} \ExternalLink
  \end{itemize}

  \vspace{2mm}

  {\bfseries Submitted:} 20 October 2020\\
  {\bfseries Published:} 24 November 2020

  \vspace{2mm}
  {\bfseries License}\\
  Authors of papers retain copyright and release the work under a Creative Commons Attribution 4.0 International License (\href{http://creativecommons.org/licenses/by/4.0/}{\color{linky}{CC-BY}}).
}

\vspace{8mm}
  
\hypertarget{summary}{%
\section{Addendum}\label{addendum}}

\texttt{pyEQUIB} is a pure Python open-source package containing several
application programming interface (API) functions that can be employed
for plasma diagnostics and abundance analysis of nebular emission lines.
This package is a Python implementation of the IDL library
\texttt{proEQUIB} \citep{ref-Danehkar:2018} that is coupled to the IDL
library \texttt{AtomNeb} \citep{ref-Danehkar:2019}. The collisional
excitation and recombination units of this package need to have the
energy levels, collision strengths, transition probabilities, and
recombination coefficients, which can be retrieved from the Python
package \texttt{AtomNeb} for \emph{Atomic Data of Ionized Nebulae}
\citep{ref-Danehkar:2020}. The API functions of this package can be used to
deduce the electron temperature, electron concentration, chemical
elements from CELs and Rls, and the interstellar extinction from the
Balmer decrements emitted from ionized gaseous nebulae. This package can
simply be used by astronomers, who are familiar with the high-level,
general-purpose programming language Python.

This package requires the Python packages \texttt{NumPy} \citep{ref-Walt:2011,ref-Harris:2020}, \texttt{SciPy} \citep{ref-Virtanen:2020}, and
\texttt{AtomNeb} \citep{ref-Danehkar:2020}. This package is released under
the GNU General Public License. The source code is publicly available on
the GitHub platform. The latest version of this package can be installed
directly from its repository on the GitHub, and its stable version from
the Python Package Index (PyPi) via \texttt{pip\ install\ pyequib} or
alternatively from the Anaconda Python package distributor via
\texttt{conda\ install\ -c\ conda-forge\ pyequib}. The online
documentation, tutorials and examples are available on its GitHub page
(\textsf{\url{https://equib.github.io/pyEQUIB/}}) and its Read the Docs documentation
page (\textsf{\url{https://pyequib.readthedocs.io/}}).

\hypertarget{acknowledgements}{%
\section{Acknowledgements}\label{acknowledgements}}

The author acknowledges the support of Research Excellence Scholarship
from Macquarie University.


\end{document}